\documentclass[aps,pre,amsmath,amsfonts,floatfix,superscriptaddress]{revtex4}
\usepackage{mathrsfs} \usepackage{graphicx} \usepackage{bbm}

\let\a=\alpha \let\b=\beta \let\g=\gamma \let\d=\delta
   
\let\l=\lambda  \let\n=\nu  \let\p=\pi
\let\s=\sigma \let\t=\tau \let\f=\varphi

\let\ee=\epsilon \let\r=\rho  \let\io=\infty

\def\ie{{i.e. }}\def\eg{{e.g. }}

\def\PP{{\cal P}}\def\EE{{\cal E}} 
\def\FF{{\cal F}} \def\HH{{\cal H}}
\def\NN{{\cal N}} \def\II{{\cal I}}
  
\def\DD{{\cal D}}\def\GG{{\cal G}}

\def\to{\rightarrow} \def\la{\left\langle} \def\ra{\right\rangle}
\def\vs{\vec\sigma}
\def\vt{\vec\tau}

\newcommand{\beq}{\begin{equation}} \newcommand{\eeq}{\end{equation}}

\begin{document}

\title{Can rare SAT formulas be easily recognized? \\
On the efficiency of message passing algorithms for $K$-SAT \\at large
  clause-to-variable ratios.} 

\author{Fabrizio Altarelli} 
\affiliation{Dipartimento di Fisica and INFM-SMC, Universit\`a di Roma ``La Sapienza'',
P.le A. Moro 2, 00185 Roma, Italy}
\author{R\'emi Monasson} 
\affiliation{Laboratoire de Physique Th\'eorique de l'\'Ecole Normale
  Sup\'erieure, 
24 Rue Lhomond, 75231 Paris Cedex 05, France}
\author{Francesco Zamponi} 
\affiliation{Laboratoire de Physique Th\'eorique de l'\'Ecole Normale
  Sup\'erieure, 
24 Rue Lhomond, 75231 Paris Cedex 05, France}

\begin{abstract}
For large clause-to-variable ratio, typical $K$-SAT instances drawn from the
uniform distribution have no solution. We argue, based on statistical 
mechanics calculations using the replica and cavity methods, that rare 
satisfiable instances from the uniform distribution are very similar 
to typical instances drawn from the so-called planted distribution, 
where instances are chosen uniformly between the ones that admit a 
given solution. It then follows, from a recent article by Feige, 
Mossel and Vilenchik, that these rare instances can be easily recognized 
(in $O(\log N)$ time and with probability close to 1) by a simple message-passing algorithm.
\end{abstract}

\maketitle

\section{Introduction}

A lot of efforts have recently been devoted to the investigation of the
computational complexity of hard computational problems under input
model distributions. One popular case is the $K$-Satisfiability
($K$-SAT) problem with uniform distribution where
clauses are picked up uniformly at random from the set of $K$-clauses 
over $N$ Boolean variables \cite{MI92}.
It is widely believed that there exists a phase transition when 
the number of clauses, $M$, and of variables, $N$, go to infinity at
fixed ratio $\alpha$. Instances with ratio $\alpha$ smaller than some
critical value $\alpha _c(K)$ typically admit a solution, while instances
with ratio $\alpha > \alpha _c (K)$ are almost surely
not satisfiable. Rigorous studies combined with
statistical physics methods have produced bounds and estimates to the
value of $\alpha _c(K)$ and conjectured the existence of a rich
structure of the space of solutions in the satisfiable phase 
\cite{FR99,DU01,AC05,MZ97,BI00,ME02,ME06}.  

If those results are certainly interesting from the random graph theory
point of view their relevance to computer science is a matter of
debate. The major concern is that they are highly specific
to one particular distribution of instances, with no obvious
theoretical generality or usefulness for practical applications where
instances are highly structured or extracted from unknown distributions. 
Recently, however,  a strong motivation to the study of random $K$-SAT
in computer science was pointed out by Feige \cite{Fe02}. Under the
assumption that 3-SAT is hard on average under uniform 
distribution Feige proved some worst-case hardness of 
approximation results for many different problems such as min bisection,
dense $K$-subgraph, max bipartite clique, etc... The average-case
hardness hypothesis can be informally stated as : 
{\em there is no fast algorithm capable of 
recognizing every satisfiable instance and most unsatisfiable
instances for arbitrarily large (but bounded when $N\to \infty$) ratio
$\alpha$}.  

In the present work we point out that a similar but stronger
hypothesis, with {\em every} replaced by {\em with probability $p$}, is wrong
whatever $p< 1$. For large $\alpha$
(well above $\alpha_c$) typical instances of the uniform
distribution have no solution. We argue, based on statistical
mechanics calculations using the replica and cavity methods, 
that rare satisfiable instances from the uniform distribution
are very similar to typical
instances drawn from the so-called planted distribution, where 
instances are chosen uniformly between the ones that admit a given 
solution. Our result then follows from a recent article by
Feige, Mossel and Vilenchik who showed that, for $K$-SAT with the
planted distribution, a simple message-passing algorithm is able to find the
solution with probability $1-e^{-O(\alpha)}$ in polynomial time \cite{FMV06}. 

\section{Definitions}
\label{sec:definizioni}

We consider random $K$-SAT instances (formulas) with $N$ Boolean variables and $M=\a N$
clauses. A clause has the form $C_i(X) = y_{i_1} \vee y_{i_2} \vee \cdots \vee
y_{i_K}$, where $y \in \{x,\bar x \}$ represents the variable or its negation; 
the $K$-SAT problem consists in finding an assignment $X$ such that 
$\wedge_j C_j(X) =$TRUE. Sometimes we will specialize to the case $k=3$ for
simplicity.
We will consider the following distributions over the formulas $F$
\cite{FMV06}:
\begin{itemize}
\item the uniform distribution $\PP_{unif}[F]$ over all possible formulas
  with $N$ variables and $M=\a N$ clauses made of $K$
  literals (variables or their negations) corresponding to {\it different} variables;
\item the distribution $\PP_{sat}[F]$ obtained from the distribution
  above by conditioning to satisfiability. In other words,
  $\PP_{sat}[F]$ gives {\it uniform} probability to all satisfiable
  formulas and zero to the others. 
\item the {\it planted} distribution $\PP_{plant}[F]$ which is
  constructed as follows: first one extracts with uniform probability
  one configuration $X$ of the $N$ variables, and then extracts with
  uniform probability one formula among the ones that admit $X$ as a
  solution. Non-uniform variants of $\PP_{plant}$ were studied in
  \cite{BA02}.  
\end{itemize}

The number of formulas that have a given solution $X$ is
independent of $X$ for symmetry reasons, $\NN_f[X]=\NN_f = [ {N
\choose K} (2^K-1)]^M $. Define
$\chi[F;X]=1$ when $X$ is a solution to $F$ and $\chi[F;X]=0$
otherwise, and  $\NN_s[F]$ the number of solutions to $F$. We
have \beq\label{planted} \PP_{plant}[F] = \frac{1}{2^N} \sum_X
\frac{\chi[F;X]}{\NN_f[X]} = \frac{\NN_s[F]}{2^N \NN_f} \ ; \eeq thus
$\PP_{plant}$ is {\it not} uniform over the satisfiable instances, but
is proportional to the number of solutions to a given formula.

In \cite{Fe02} the following hardness hypothesis was introduced 
for formulas drawn from the uniform distribution $\PP_{unif}$,

\vskip5pt

{\sc Hypothesis 1:} {\it Even if $\a$ is arbitrarily large (but
  independent of $N$), there is no polynomial time algorithm that on
  most 3-SAT formulas outputs UNSAT, and always outputs SAT on a 3-SAT
  formula that is satisfiable}.
\vskip5pt \noindent and used to derive  hardness of approximation results
for various computational problems.  A stronger form of
hypothesis 1 is obtained by replacing {\em never} with {\em
 with probability $p$} (with respect to the uniform distribution over the
formulas and possibly to some randomness built in the algorithm):
\vskip5pt

{\sc Hypothesis $1_p$:} {\it Even if $\a$ is arbitrarily large (but
  independent of $N$), there is no polynomial time algorithm that on
  most 3-SAT formulas outputs UNSAT, and outputs SAT with probability
  $p$ on a 3-SAT formula that is satisfiable}.

\vskip5pt We want to present some arguments supporting the idea that
$1_p$ is false for any $p<1$.  Indeed, in \cite{FMV06} it has been
shown that, if the formulas are drawn with probability $\PP_{plant}$,
then a solution is found in polynomial time with probability
$1-e^{-O(\a)}$ by a message-passing algorithm, called Warning
Propagation (WP). WP is a simplified version of the zero-temperature Belief 
Propagation procedure, see \cite{FMV06,MCKAY} for a presentation. It is
important 
to notice that WP is a constructive algorithm: when it declares a formula 
to be satisfiable, it provides a solution. This means that it never outputs 
SAT on a formula which is unsatisfiable. On the other hand,
if the algorithm has not found a solution after
a given number of iterations (which depends on $N$, see below), we declare the 
output to be UNSAT.

It is natural (and was already suggested in \cite{FMV06}) to try to extend
this result to formulas drawn from the distribution $\PP_{sat}$.  The
main ingredients that are needed in the proof of \cite{FMV06} are the
following:
\begin{enumerate}
\item at large $\a$, formulas drawn from $\PP_{plant}[F]$ typically
  have a single {\it cluster} of solutions with a large {\it core}:
  namely, there is a set $\HH$ (the core) containing a fraction $
  1-e^{-O(\a)}$ of variables that have the same value in all the solutions of a
  given formula drawn from $\PP_{plant}[F]$;
\item the {\it cavity fields} (or {\it variable-to-clause messages})
  corresponding to the core variables, defined roughly as the number
  of clauses that are violated if one takes a solution to $F$ and
  changes the value of a given core variable, are $O(\a)$;
\item the cavity fields for the core variables are $O(\a)$ even if
  they are computed with respect to a random configuration (see
  \cite{FMV06} for a precise definition); this is a consequence of the
  fact that if a variable $x_i$ has value $1$ in the solutions to $F$,
  then the probability of this variable appearing as $x_i$ in a clause
  (according to $\PP_{plant}$) is bigger than the probability of it
  appearing as $\bar x_i$ (and viceversa if the variable is $0$ in the
  solutions).
\end{enumerate}
We claim that formulas drawn from $\PP_{sat}$ are very similar
to the ones drawn from $\PP_{plant}$, 
and in particular properties 1, 2 and 3 hold for them.
Moreover we will show that the relative entropy (Kullback-Leibler
divergence) of $\PP_{plant}$ with
respect to $\PP_{sat}$ is $O(Ne^{-\a})$.
In particular property 1 implies that properties of
the formulas drawn from $\PP_{sat}$ (such as the distribution of the
cavity fields) can be computed in a {\it replica symmetric}
framework. We will indeed show that this is the case as the replica
symmetric solution is stable for large $\a$ if one restricts to SAT
formulas. In this way we will compute the distribution of the cavity
fields in a solution to show that: {\it i)} only a fraction
$e^{-O(\a)}$ of the fields are zero (corresponding to non-core
variables); {\it ii)} the non-vanishing cavity fields are typically
$O(\a)$; {\it iii)} if the field corresponding to a variable $x$ is
(say) positive, then the number of clauses where the literal
$x$ appears is bigger
than the number of clauses where $\bar x$ appears, the difference
being $O(\a)$.

The validity of properties 1-3 together with the fact that the
relative entropy of $\PP_{sat}$ and $\PP_{plant}$ is small strongly
suggest that the analysis of \cite{FMV06} can be extended to
$\PP_{sat}$. Then WP will be efficient in finding
solutions for satisfiable formulas in polynomial time, with probability
close to 1 for large $\a$, thus contradicting hypothesis $1_p$ (but \emph{not} hypothesis 1).

\section{Statistical Physics Analysis of $\PP_{sat}$}

\subsection{From $\PP_{unif}$ to $\PP_{sat}$: the replica calculation}
\label{replicas}

We want to compute properties of the satisfiable formulas drawn from the
uniform distribution $\PP_{sat}[F]$ using the replica method. Following
\cite{MZ97} we introduce a cost function
\beq
E[X]=\sum_{i=1}^M \d[C_i(X);\text{FALSE}] \ ,
\eeq
where the function $\d[C_i(X);\text{FALSE}]$ is $1$ if clause $C_i$ is false
in the assignment $X$
and $0$ otherwise (\ie $E$ counts the number of violated clauses).
The replicated and disorder-averaged (\ie averaged over the distribution
$\PP_{unif}[F]$ of the formulas) partition function is
\beq
\overline{Z(\b)^n} = \overline{\left[ \sum_X e^{-\b E[X]} \right]^n}
= \overline{\left[ g_0 e^{-\b E_0} + g_1 e^{-\b E_1} + \cdots  \right]^n}
\eeq
where $E_0$ is the energy of the ground state (\ie the minimal number of
unsatisfied clauses in $F$) and $g_0$ its degeneracy.
In the limit $\b \to \io$, $n \to 0$ with fixed product $\n=n \beta$, defining
\beq
P(E_0 = N e_0) = e^{N \omega(e_0)+o(N)}
\eeq
the distribution of the ground state energy with respect to $\PP_{unif}[F]$, we have
\beq \label{integral}
\overline{Z(\b)^n} \sim \overline{ g_0^n e^{-n \b E_0}} = \int de_0 \
e^{N [\omega(e_0) -
  \n \;e_0]} + O(e^{-\b}) = e^{N \FF(\n)}
\eeq
since $g_0$ is independent of $n$ and therefore disappears for $n \to 0$. 
The function $\FF(\n)$ is defined by the saddle point condition
$\FF(\n)=\max_{e_0} [\omega(e_0) - \n e_0]$. 
We will verify later that $\FF(\n)$ is convex (for sufficiently large $\a,\n$), so that $\FF(\n)$ and
$\omega(e_0)$ 
are the Legendre transforms of each  other. 
The dominant contribution to the integral in (\ref{integral}) comes
from formulas with ground state energy density $e_0$ given by the
equation ${e_0} (\n) = -\partial_\nu \FF(\n)$. As we will see from the calculation of $\FF(\n)$, for large $\n$ we have $e_0(\n) \sim e^{-\n}$, so that to have $e_0 (\nu)=0$ we have to take the limit $\n \to \io$.
By imposing this limit we implement the constraint $e_0 = 0$, and obtain 
information on $\PP_{sat}[F]$. Note that we cannot implement the exact
constraint of satisfiability, $E_0 =0$, but only $\lim_{N\to\io} E_0/N =
0$, as usual in most statistical mechanics computations. All our results are
then affected by corrections vanishing only for $N\to\io$.

In a replica symmetric framework the free energy $\FF (\n)$ is
obtained by maximizing a functional $\FF[R(z),\n]$ 
over a functional order parameter $R(z)$
(Appendix~\ref{app:Freplica}). This order parameter is the probability
distribution of a random field $z_i$ acting on each variable. The latter is
 the difference between
the minimal number of violated clauses when the variable $x_i$ is set to
TRUE and the same quantity for $x_i=$FALSE. The distribution 
$R(z)$ is determined by the saddle point equations $\d \FF[R(z),\n] / \d R(z)
= 0$, which admit a solution in 
which the fields $z$ are integer valued, as expected, 
\begin{equation}
R(z) = \sum_{n=-\infty}^{+\infty} r_n \ \delta(z-n). \ \label{sommarn}
\end{equation}
The coefficients $r_n$ are obtained by substituting this expression
into the saddle point equations and 
solving them (Appendix~\ref{app:saddle}). 
In the limit $\n\to\io$ the saddle-point equations give a self-consistency
equation for $\r_0 = \lim_{\nu\to\io} r_0$:
\beq\label{rho0text}
\begin{split}
\r_0 &= \frac{1}{2e^{\GG(\r_0)}-1} \ , \\
\GG(\r_0) &=  \frac{\a K}2
\frac{\left(\frac{1-\r_0}{2}\right)^{K-1}}{1-\left(\frac{1-\r_0}{2}\right)^K
} \ , \\
\end{split}\eeq
while all the other coefficients are given by a Poissonian distribution
\beq \label{Poisson0}
\r_n = \lim_{\nu\to\io} r_n= \frac{\GG(\r_0)^{|n|}}{|n|!}
\frac{1}{2 e^{\GG(\r_0)} - 1} \ .
\eeq 
Using the expressions above, it is possible to show that 
the ground state energy vanishes exponentially for $\nu\to\io$,
$\overline{e_0}(\n) = - \partial_\n \mathcal
F(\nu) = C e^{-\n}$ (Appendix~\ref{app:energy}).

We shall be interested in the value of $\omega(e_0=0)$ which is related to the
probability of a formula extracted from $\PP_{unif}[F]$ being satisfiable;
setting from (\ref{egscompleta}) $e_0 \sim e^{-\n}$ we obtain that 
$\omega(e_0(\n)) = \FF(\n) + \n e_0(\n) \Rightarrow \omega(0) = \FF(\io) +
O(\n e^{-\n})$ 
and using the result for $\FF(\n)$ given in Appendix~\ref{app:free}, we get
\beq 
\omega(0) = \FF(\io) = \log[2 e^{\GG} -1] - \frac{2 \GG
e^{\GG}}{ 2e^{\GG}-1} + \a \log\left[ 1 -
\left(\frac{1-\r_0}{2}\right)^K \right] \ .
\eeq
where $\GG \equiv \GG(\r_0)$.

To conclude this section, let us discuss the stability of the replica symmetric
solution for $\n\to\io$, \ie for satisfiable formulas. The eigenvalues
of the stability matrix around the saddle-point are calculated in 
Appendix~\ref{app:stability}. We show that all the eigenvalues are
negative for large enough $\a$ and $\n\to\io$.
This implies that the replica symmetric solution is {\it locally}
stable,
but does not exclude the existence of a first order transition to a
different solution.  In Appendix~\ref{app:nonint} we show that if one
considers a function $R(z)$
different from (\ref{sommarn}) that allows also non-integer values of the
fields, the weights of the non-integer fields vanish for $\a > \a_s$ 
(for some constant $\a_s$ which depends on $K$)
and one gets back Eq.~(\ref{sommarn}). This result rules out the
existence of a first-order phase
transition in the replica symmetric subspace. 
To exclude the possibility of a first order transition with replica
symmetry breaking 
one should perform the full 1RSB computation, that we leave for future work.


\subsection{Interpretation of the self-consistency equations for the
  field distribution}
\label{cavity}

Self-consistency equations (\ref{rho0text}) can be found back within
the cavity method. Consider a formula over $N-1$ variables, and add a new
variable $x$, by connecting it to the others $N-1$ variables 
through $\ell_+$ clauses where $x$ enters, and $\ell_-$ clauses where 
$\bar x$ enters. We assume that $\ell_+,\ell_-$ are independent
Poissonian variables, with probabilities
\beq
p_L(\ell_\pm) = \frac{(\a' K/2)^{\ell_\pm}}{\ell_\pm!} e^{-\a' K/2} \ \label{occurrences}
\eeq
where $\alpha'$ is some constant to be determined later.
The signs of the other $K-1$ variables in each clause are chosen
uniformly at random. Then the probability that a new clause 
constrains the value of $x$ (the clause sends a
message to $x$ in WP language \cite{FMV06,MCKAY}) is equal to
\beq\label{qdef}
q = \left(\frac{1-\r_0}{2}\right)^{K-1} \ ,
\eeq
as $(1-\r_0)/2$ is the probability that the field on an `old' variable
due to the existing formula is in
contradiction with its sign in the new clause: \eg if $C = x \vee x_1 \vee \cdots \vee x_{K-1}$
this is the probability that all the fields on $x_1,\cdots,x_{K-1}$ are
negative, so that $C$ sends a message (``be 1'') to $x$.
Let us call $m_+,m_-$ the numbers of clauses containing, respectively,
$x,\bar x$ and sending messages to the new variable. These are
stochastic independent variables with probabilities 
\beq
p_M(m_\pm) = \sum _{\ell=m_\pm} ^\infty p_L (\ell) \; {\ell \choose m_\pm}
\; q^{m_\pm}\; (1-q)^{\ell - m_\pm} = \frac{(\a' K q/2)^{m_\pm}}
{m_\pm!}\; e^{-\a' Kq/2} \ . \label{messaggi}
\eeq
We will also need later on the weighted distribution of $m_\pm$, where the
weight is the number of clauses of type $\pm$, 
\beq
p^{(w)}_M(m_\pm) = \sum _{\ell=m_\pm} ^\infty \ell\; p_L (\ell) \; 
{\ell \choose m_\pm}
\; q^{m_\pm}\; (1-q)^{\ell - m_\pm} = p_M(m_\pm)\times \left( m_\pm +
\frac{\a' K}2 (1-q) \right) \ . 
\eeq
Given $m_+,m_-$ the best value for the variable $x$ is TRUE if $m_+ >
m_-$ and FALSE otherwise. The minimal number of violated
clauses in the formula therefore increases by $E = \min
(m_-,m_+)$. The field acting on the variable $x$ is the difference
between the number of violated clauses when $x$ is TRUE and when it is
FALSE, $z=m_+-m_-$. In particular the formula keeps being satisfiable
upon inclusion of the new clauses if $m_-$ or $m_+$ is equal to zero.
The joint probability of the increase in energy and of the field reads
\beq \label{h1}
P(E,z) = \sum_{m_+=0}^\infty p_M(m_+)\sum _{m_-=0}^\infty  p_M(m_-) \;
\delta _{E,\min (m_-,m_+)}\; \delta _{z,(m_+-m_-)}\ . 
\eeq
Introducing the chemical potential $\nu$ of Section \ref{replicas} we
weight each formula with a factor $\exp(-\nu \, E)$. The probability
that the new variable is subjected to a field equal to $z=n$ is thus
\beq \label{h2}
r_n (\nu) = \frac{ \sum _{E\ge 0} P(E,n) \, e^{-\n \, E}}
{ \sum _{E\ge 0} \sum _m P(E,m) \, e^{-\n \, E}}
\eeq
where the denominator takes into account the proper reweighting of all
formulas at fixed chemical potential $\n$.

We now restrict to the case of satisfiable formulas $\n\to\infty$.
From (\ref{h1}),(\ref{h2}) we obtain the probability of having zero
field on $x$ given that the formula is
satisfiable, 
\beq\label{r0cavity}
\r_0 = \lim _{\nu \to \infty} r_0 (\n) =  \frac{ P(0,0) }
{ \sum _m P(0,m) } = 
\frac{1}{2  e^{\frac{\a' K q }{2}} - 1} \ .
\eeq
We have now to take into account the fact that we want each variable to appear
in $\alpha K$ clauses on average. Even if $\ell_\pm$ are distributed according
to a Poissonian with average $\a' K/2$, the fact that
when we add the clauses we must discard the possibilities in which the
variable $x$ receives contradictory messages makes the average number of
clauses we add smaller than $\a' K$. It is easy to check, using (\ref{qdef}) 
and (\ref{r0cavity}), that
\beq\label{ellemedio}
\langle \ell_+ + \ell_- \rangle =
\frac{\sum_{m_+,m_-} \big( p^{(w)}_M(m_+)\, p_M(m_-) +
 p_M(m_+)\, p^{(w)} _M(m_-) \big) \; \delta _{0,\min(m_-,m_+)}}
{  \sum _m P(0,m)}
= \a' K \left[ 1 - \left(\frac{1-\r_0}2\right)^K \right] \ .
\eeq
Then if we want $\langle \ell_+ + \ell_- \rangle = \a K$ we have to
renormalize $\a'$ as
\beq
\a' = \frac{\a}{1 - \left(\frac{1-\r_0}{2}\right)^{K}} \ ,
\eeq
and substituting in (\ref{r0cavity}) we get back Eq.~(\ref{rho0text}) 
as $\frac{\a' K q }{2} = \GG(\r_0)$.

The generating function $G(x)$ of the distribution of variable occurrences 
$\ell = \ell_+ + \ell_-$ can be easily
computed by adding a weight $x^\ell$ in (\ref{messaggi}) and summing over
$m_\pm$ with the constraint $\min(m_+,m_-)=0$; after a correct
normalization one obtains
\beq
G(x) = e^{\a' K (x-1) (1-q)} \frac{2e^{\a' K x q/2} - 1}{2e^{\a' K q/2} - 1} \ ,
\eeq
that generates the difference of two Poissonians distribution with different parameters.
This distribution differs from the normal Poissonian distribution of
occurrences. However the difference is exponentially small in $\a$ for all
values of $x$, indeed for large $\a$ we get (recalling that $\r_0 \to 0$)
\beq
G(x) = e^{\a' K (x-1) (1-q/2)} + e^{-O(\a)} = e^{\a K (x-1)} + e^{-O(\a)} \ ,
\eeq
which is the generating function of a Poissonian with parameter $\a K$,
consistently with (\ref{ellemedio}). 
The fact that the distribution is Poissonian implies that the
cavity fields have the same distribution of the true fields; the latter has been
obtained from the replica method in Section \ref{replicas}.

\section{Comparison of $\PP_{sat}$ and $\PP_{plant}$ at large ratio $\alpha$}

For large $\a$ the solution to (\ref{rho0text}) is well approximated by
\beq 
\r_0 = \frac{1}{2 e^{\g} - 1} \ , \hskip30pt \g \equiv \GG(0) =
\frac{\a K}{2^K -1} \ .  
\eeq 
The distribution of the fields becomes
\beq
\label{Poisson} 
\r_n = \frac{\g^{|n|}}{|n|!}
\frac{1}{2 e^{\g} - 1} \ .
\eeq
This result implies that the solutions of formulas 
extracted from $\PP_{sat}[F]$ are very similar to each other. They differ 
by a fraction 
$e^{-O(\a)}$ of the variables only (this is in fact the
weight $r_0$ of the field $z=0$). The remaining fields $z\neq 0$ have
typically values $O(\g) = O(\a)$, so that the variables in the core
(with the same assignments in all the solutions) have
strong cavity fields pointing to their correct assignments. 
Moreover there is
a correlation between cavity fields (or equivalently, between values of the variable in the
solutions) and occurrence of the variables in the formula, as
discussed in the next section.
We will in addition show that the results (\ref{Poisson}), 
coincide with the ones obtained from the planted
distribution, thus indicating that the two distributions coincide with errors
$e^{-O(\a)}$; 
indeed we will compute the relative (extensive) entropy of the distribution and show
that it is of the order of $N e^{-O(\a)}$.

\subsection{Distribution of the fields}

To show that the Poissonian distribution (\ref{Poisson}) is the 
same distribution
that is obtained from the planted distribution recall that $\r_n$ is the
probability of violating $|n|$ clauses when a variable $x_i$ is flipped
from the correct value it has in the ground state to the opposite
value. The sign of $n$ is positive if $x_i=$TRUE in the solution and negative otherwise.

The planted distribution is constructed by extracting at random a
configuration $X$, called root, and giving the same probability to 
all the choices of
the clauses for which $X$ is a solution. Then, if we choose at
random a set of $K$ indices $(i_1,\cdots,i_K)$, and consider all the
possible $2^K$ equations we can construct with these indices, we see
that only one of the choices is not allowed (e.g. if the configuration
$X$ is such that $(x_{i_1},\cdots,x_{i_K}) = (1,\cdots,1)$, only the choice
$\bar x_{i_1} \vee \cdots \vee \bar x_{i_K}$ is not allowed).

The probability of violating $|n|$ clauses can be computed as follows.
For simplicity we choose the root $X = (1,\cdots,1)$. Then we want to know how
many clauses are violated when we flip one variable, e.g. $x_1 \to 0$.
The clauses that are violated have the form 
\beq\label{falsa} 
x_1 \vee \bar x_{i_2} \vee \cdots \vee \bar x_{i_K} 
\eeq 
The probability $p(|n|)$ that $|n|$ such clauses appear in a formula is a
binomial distribution with parameter $p$ given by the product of the probability that 
the variable $x_1$ appears in the clause, which is $K/N$, times the probability that
the signs of the variables in the clause make it unsatisfied by the flipped assignment, 
which is $1/(2^K-1)$:

\beq p =
\frac{K}{N} \frac{1}{2^K-1} \ . \eeq The number of equations is $M=\a N$
and they are independent so the probability of violating $|n|$ equations is
\beq
p(|n|) = \binom{M}{|n|} p^{|n|} (1-p)^{M-|n|} \sim \frac{\g^{|n|}}{|n|!} e^{-\g} \ .
\eeq
In a generic root $X$ almost half of the
variables are TRUE, giving rise to a positive field, while 
the other half are FALSE and correspond to negative fields; then we have
\beq
\r_{plant}(n) = e^{-\g} \d_{n0} + \frac{\g^{|n|}}{2\, |n|!}e^{-\g} (1-\d_{n0}) \ .
\eeq
This distribution differ from (\ref{Poisson}) by $e^{-O(\a)}$.

\subsection{Correlation between fields and occurrences of the negations}
\label{correl}

Most variables are typically subject to strong fields, of
the order of $\alpha$ in absolute value, in ground state assignments. 
We now show that the sign of the field $z$ associated to a variable, say, $x$,
is strongly correlated to the numbers of occurrences of literals $x$
and $\bar x$ in the formula.
 
Consider the cavity derivation of the self-consistent equations for
the fields exposed in Section \ref{cavity}. Suppose $z>0$, and require the
formula to be satisfiable ($\n\to\infty$). Then the number of
messages coming from $\pm$-type clauses are $m_-=0,m_+\ge 1$. 
We define the average values of the number of $\pm$-type clauses, 
$\la \ell_\pm \ra_{z>0}$, as follows:
\beq
\la \ell_\pm \ra_{z>0} = \frac{\sum_{m_+\ge 1} p^{(w)} _M (m_+) \,
  p_M(0)} {\sum_{m_+\ge 1} p_M (m_+) \,
  p_M(0)}  \ .
\eeq
We get
\beq\begin{split}
&\la \ell_+ \ra_{z>0} = \frac{\a' K}{2} \left[ \frac{1-(1-q)
    e^{-\GG}}{1-e^{-\GG}} \right] = \frac{\a K}{2} 
\frac{1}{1-2^{-K}} + e^{-O(\a)}
 \ , \\
&\la \ell_- \ra_{z>0} = \frac{\a' K}2 (1-q) = \frac{\a K}{2} 
\frac{1-2^{-(K-1)}}{1-2^{-K}}+ e^{-O(\a)} \ ,
\end{split}
\eeq
and finally
\beq\label{frazpos}
\frac{\la \ell_+ \ra_{z>0} - \la \ell_- \ra_{z>0}}{\la \ell_+ \ra_{z>0} + \la
  \ell_- \ra_{z>0}}
=\frac1{2^K-1} + e^{-O(\a)} \ .
\eeq

The average value of the bias between the numbers of positive and
negative occurrences found for $\PP_{sat}$ coincides with its
counterpart for $\PP_{plant}$ at large ratios $\a$.
Consider again the planted distribution with respect to $X=
 (1,\cdots,1)$. It is easy to show that variables occur more
frequently non-negated \cite{BA02}. 
Indeed, if $x$ enters a given clause $C$,
 there are $2^K-1$ possible assignments of the negations, among which
$2^{K-1}$ contain $x$ and $2^{K-1}-1$ contain $\bar x$ (because the assignment
 in which all variables are negated is forbidden). Then it is clear that
\beq
\frac{\la \ell_+ \ra_{plant} - \la \ell_- \ra_{plant}}
{\la \ell_+ \ra_{plant} + \la \ell_- \ra_{plant}} = \frac{1}{2^K-1} \ ,
\eeq
as in (\ref{frazpos}).

\subsection{Relative entropy}
\label{relative}

The relative entropy of $\PP_{plant}$ to $\PP_{sat}$ is given, using
(\ref{planted}), by
\beq
\s =- \sum_F \PP_{sat}[F] \log \frac{\PP_{plant}[F]}{\PP_{sat}[F]} =
-\log \NN_{sat} + N \log 2 + \log \NN_f -
 \sum_F \PP_{sat}(F) \log \NN_s[F] \label{dkl}
\eeq
where $\NN_{sat}$ denotes the number of satisfiable formulas.
We have
\beq
\NN_{sat} =  e^{N \omega(0)} \times \NN \qquad \mbox{where}\qquad \NN=
\left[ \binom{N}{K}2^K \right]^M
\eeq 
is the total number of formulas. 
Using $\r_0 \sim e^{-\g}/2$, $\GG(\r_0) \sim \g + O(e^{-\g})$,
$\r_0 e^{\GG(\r_0)} \sim 1/2
+ O(e^{-\g})$, we get
\beq\label{om0approx}
\omega(0) \sim \log 2 + \a \log \left( 1 -\frac1{2^K} \right) + \frac12 \g
e^{-\g} + O(e^{-\g}) \ .
\eeq
The value of the number of formulas sharing a common root, $\NN_f$,
was given in Section \ref{sec:definizioni}. The last term in (\ref{dkl}) 
represents the average entropy of satisfiable
formulas. It is bounded from above  by $N \rho_0 \log 2 \sim N e^{-\g}$,
because $\r_0$ is an upper bound to the fraction of variables that can
change values from solution to solution (inside the unique cluster).

Gathering all contributions we get the following expression for the
relative entropy, valid for large ratios $\a$,
\beq
\s = \frac12 N\g\; e^{-\g} + O(N e^{-\g}) \ .
\eeq
Hence $\sigma$ is extensive in $N$, and decreases exponentially with
$\alpha$.

\section{Finite energy results}

The previous results extend to formulas having a small minimal
fraction of unsatisfied clauses. This point is interesting since
the relationship between approximation hardness and average-case
complexity can be deduced from a weaker form of hypothesis
1~\cite{Fe02},
 \vskip5pt

{\sc Hypothesis 2} {\it For every fixed $\ee > 0$, for $\a$
arbitrarily large (but independent of $N$), there is no polynomial
time algorithm that on most 3-SAT formulas outputs typical, and never
outputs typical on 3-SAT formulas with $(1-\ee)M$ satisfiable clauses}.

 \vskip5pt
If we choose $\nu$ to be a large, finite number we find from the above
replica calculation that the ground state energy (\ref{egscompleta}) 
dominating the integral (\ref{integral}) becomes 
\beq\label{EGSapprox}
e_0(\n) \sim \frac{\g}{K}[1 + O(\g^2 e^{-\g})]\;  e^{-\n}  
\eeq 
for large $\a$. 
As in the $\nu\to\infty$ case, most cavity fields are non zero and typically of
the order of $\alpha$. In addition, using the calculation of Section
\ref{correl}, we can extend the calculation of the average
difference between the number of $\pm$ occurrences of a variable with
positive field, see Section \ref{correl}, to the case of large but
finite $\n$ with the result,
\beq\label{frazpos1}
\frac{\la \ell_+ \ra_{z>0} - \la \ell_- \ra_{z>0}}{\la \ell_+ \ra_{z>0} + \la
  \ell_- \ra_{z>0}}
=\frac1{2^K-1} - \frac{\a K} {2 (2^{K}-1)^2} \; e^{-\nu} + O(\a^{-1}) \ ,
\eeq
to first order in $e^{-\n}$. Eliminating $\nu$ between (\ref{EGSapprox})
and (\ref{frazpos1}) we obtain
\beq\label{frazpos2}
\frac{\la \ell_+ \ra_{z>0} - \la \ell_- \ra_{z>0}}{\la \ell_+ \ra_{z>0} + \la
  \ell_- \ra_{z>0}}
=\frac1{2^K-1} \left[ 1- e\; K \, 2^{K}\, \left( \frac 12 - \frac
  1{2^{K+1}-2} - \frac {2^K}{\alpha\, K}\right) \right] 
\eeq
to first order in the ground state energy density, $e$.
This suggests that also
formulas that are not exactly satisfiable but have few violated
clauses ($e \ll 2^{-K+1}/K$) can be detected by the WP algorithm.
A consequence is that
a weaker version of Hypothesis 2 in which ``never'' is replaced with
``with probability $p$'' should also be false for any $p>0$.

We have checked the validity of this prediction on the following
distribution of formulas, referred to as $\PP^{(E)}_{plant}$
hereafter. Pick up uniformly
at random a configuration $X$ of the variables, and choose $M$ times
independently a set of $K$ indices uniformly over the $\binom{N}{K}$ possible
ones to build $M$ clauses.
For the first $E$ clauses, the negations of the variables are chosen such that
the clause is violated by $X$ (there is only one such assignment), while for
the remaining $M-E$ clauses the negations are chosen such that the clause is
satisfied in $X$ (there are $2^K-1$ such assignments and we choose one of them
at random, as in the planted distribution).
A simple calculation similar to the one of Section \ref{relative}
shows that the relative entropy, $\sigma (e=\frac EN)$, of 
$\PP^{(E)}_{plant}$ and $\PP_{unif}$ constrained to
formulas with ground state energy $E$ is $\sigma (e) = N [ e^{-O(\a)} (1+O(e))+ 
O(e^2)]$. Thus both distributions are similar in the large
$\alpha$ limit, at least for small enough $e$.

From a numerical point of view
we extracted $3$-SAT formulas from $\PP^{(E)}_{plant}$ with $N=200$
variables, $M=2000$ (\ie $\a=10$) and studied the convergence of the WP
algorithm as a function of $E$. When the algorithm converges, it returns a
partial assignment of the variables \cite{FMV06}, 
the unassigned variables having a zero cavity field.
Without entering into a detailed numerical
investigation, we roughly observed that for $E < 10$ the algorithm behaves
essentially as for $E=0$: it converges after few iterations,
and in the returned partial assignment most of the
variables ($\sim 197 \sim N (1-e^{-\g})$) have the same value they have in the
reference configuration $X$, and the rest of the variables are
unassigned. After optimization over the unassigned variables, the
energy of the resulting configuration differs from $E$ by $\sim N e^{-\g}\sim 3$
at most. Note that $E =10$ corresponds to $e=10/200=0.05$ and is 
compatible with 
the value $e_c \sim 2^{-K+1}/K\sim 0.08$ we found above
(\ref{frazpos2}) (there are corrections
proportional to $N^{-1/2}$). Above $E \sim 15$ the
probability of convergence decreases, and the number of
unassigned variables increases, but when the algorithm converges and one 
optimizes over the unassigned variables, the resulting configuration has an 
energy close to $E$ by $\sim 3$. Above $E \sim 50$ the algorithm
almost never converges. Finally,
it is interesting to observe that, when the algorithm converges its does
so after $\sim \log N \sim 6$ iterations, 
as predicted in \cite{FMV06} for $E=0$, independent of the value of $E$.
If convergence is not attained after $\sim 10$ iterations, it is very 
likely that the algorithm will not converge in the following iterations. 
This allows one to put a cut-off on the number of required
iterations {\it a priori}.

\section{Conclusions}

The present work supports the claim that satisfiable formulas from the
uniform distribution can be recognized in polynomial time with
probability close to unity provided the ratio of clause-per-variable
is made large enough. In other words WP should be efficient to solve
the random $K$-SAT problem at large $\alpha$. This claim comes from
the closeness of the two distributions $\PP_{sat}$ and $\PP_{plant}$
for large but finite ratios $\a$. More precisely both distributions
produce formulas that {\it (1)} have a single cluster of solutions, in
which {\it (2)} a large fraction $1-e^{-O(\a)}$ of variables are
strongly constrained (they have the same value in all solutions and a
cavity field $O(\a)$) and a small fraction $e^{-O(\a)}$ is free to
change its value (zero cavity field). Moreover, {\it (3)} as shown in
section~\ref{correl}, a positively constrained variable $x$ (\ie TRUE
in all the solutions) is very likely to appear more times as $x$ than
as $\bar x$ in the formula.  The efficiency of WP on $\PP_{plant}$
relies on these properties, and therefore extends to $\PP_{sat}$, then
to $\PP_{unif}$ once a cut-off (of the order of $\log N$) is imposed
on the number of iterations. Furthermore these results extend to the
case of a small but finite energy. Formulas with a minimal fraction of
unsatisfied clauses larger than zero but much smaller than $2^{-K}$
(the typical value at large $\alpha$) can be recognized with large
probability by WP in polynomial time.

Yet the above findings are somewhat unsatisfactory for the
following reason. It is easy to repeat the statistical mechanics
calculations presented here for other Boolean functions expressing the
truth values of clauses from the variables e.g. the XORSAT model 
\cite{xorsat}. The outcome is that at large ratios properties {\it (1)}
and {\it (2)} hold quite generally but property {\it (3)} does not.
Hence while from a probabilistic point of view the solution spaces of
satisfiable SAT and XORSAT formulas far above the threshold are
similar, they are not from an algorithmic point of view. More precisely
WP cannot find out whether a XORSAT formula is typical (and has a
minimal fraction of unsatisfiable clauses close to $\frac 12$) or
exceptional (minimal fraction $e\ll \frac 12$). It would be thus
interesting to devise an algorithm capable of performing this
task. What implications this would have on hypothesis 2 remains to be
clarified too.

\acknowledgments

We wish to thank Giorgio Parisi for many discussions and for a
careful reading of the manuscript, and Uriel Feige for useful comments.
This work has been supported by the
EU Research Training Network STIPCO (HPRN-CT-2002-00319).


\appendix

\section{Replicated free energy}
\label{app:Freplica}

Here we sketch the derivation of the replicated free energy following
\cite{MZ97}.
The partition function can be written as $Z(\b) = \sum_X \prod_{i=1}^M e_i(X)$,
where $e_i(X) = 1$ if the clause $C_i=$TRUE in configuration $X$ and $e^{-\b}$
otherwise. Then
\beq
\overline{Z(\b)^n} = \overline{\sum_{X_1\cdots X_n} \prod_{i=1}^M e_i(X_1)
  \cdots e_i(X_n) } = \sum_{X_1\cdots X_n} \prod_{i=1}^M \overline{e_i(X_1)
  \cdots e_i(X_n)} = \sum_{X_1\cdots X_n} \left[ 
 \overline{e(X_1)
  \cdots e(X_n) } \right]^M \ ,
\eeq
as the clauses are all chosen independently and with the same probability
distribution. It is convenient to represent the variables $x_i$ as spins, 
i.e. $x_i =0 \leftrightarrow \s_i = -1$ and $x_i = 1 \leftrightarrow \s_i =
1$; then $\s_i^a$ denotes the value of the spin at site $i$ for replica $a$, 
$\vec \s_i$ is the $n$-component vector of the replicas of site $i$, 
$\underline{\s^a}$ is the $N$-component vector of the configuration of replica
$a$, and $\underline{\vs}$ is the full replicated configuration.
Then we can compute
\beq
\overline{e(\underline{\s^1}) \cdots e(\underline{\s^n})} = \frac{1}{{N \choose K}} \sum_{i_1 < \cdots < i_K}^{1,N} 
\frac{1}{2^K} \sum_{q_1 \cdots q_K}^{-1,1} \prod_{a=1}^n \left\{
1 + (e^{-\b}-1) \prod_{\ell=1}^K \d[\s^a_{i_\ell},q_\ell] \right\} \ ,
\eeq
where the variables $q_\ell$ correspond to the random choice of the negation
in the clause $C$ ($q_\ell = 1 $ means that the variable $x_{i_\ell}$ is
negated in $C$). To leading order in $N$ we can neglect the constraint that
all the $i$'s have to be different, and
replace $\binom{N}{K}^{-1} \sum_{i_1 < \cdots < i_K}^{1,N}$ with
$N^{-K} \sum_{i_1, \cdots, i_K}^{1,N}$.

Introducing the order parameter
\beq
\r(\vec \t|\underline{\vs})=\frac{1}N \sum_{i=1}^N \prod_{a=1}^n \d[\t^a,\s^a_i] \ ,
\eeq
that counts the number of sites $i$ such that $\vs_i=\vec \t$, we can write
\beq
\overline{e(\underline{\s^1}) \cdots e(\underline{\s^n})} = \sum_{\vec \t_1\cdots \vec \t_K}
\r(\vec \t_1|\underline{\vs})\cdots \r(\vec \t_K|\underline{\vs}) \EE(\vec \t_1,\cdots,\vec \t_K) \ ,
\eeq
where
\beq\label{epsilonG}
\EE(\vec \t_1,\cdots,\vec \t_K) = \frac1{2^K} \sum_{q_1\cdots q_K}^{-1,1}
\prod_{a=1}^n \left[ 1 + (e^{-\b}-1) \prod_{\ell=1}^K \d[\t^a_\ell,q_\ell]
\right] \ .
\eeq
Finally we write
\beq\begin{split}
\overline{Z(\b)^n} &= \sum_{\underline{\vs}} e^{M \log \sum_{\vec \t_1\cdots \vec \t_K}
\r(\vec \t_1|\underline{\vs})\cdots \r(\vec \t_K|\underline{\vs}) \EE(\vec \t_1,\cdots,\vec \t_K)} \\
&= \int_0^1 \d c(\vec \t) \ e^{N \a \log \sum_{\vec \t_1\cdots \vec \t_K}
c(\vec \t_1)\cdots c(\vec \t_K) \EE(\vec \t_1,\cdots,\vec \t_K)}
\sum_{\underline{\vs}} \prod_{\vec \t} \d[c(\vec \t)-\r(\vec \t|\underline{\vs})] \ ,
\end{split}\eeq
and observing that
\beq
\sum_{\underline{\vs}} \prod_{\vec \t} \d[c(\vec \t)-\r(\vec \t|\underline{\vs})] = 
\frac{N!}{\prod_{\vec \t} [N c(\vec \t)]!} \sim e^{-N \sum_{\vec \t} c(\vec \t) \log c(\vec \t)} \ ,
\eeq
we finally obtain
\beq\label{repZ}\begin{split}
&\overline{Z^n[\b]} = \int_0^1 dc(\vec \t) e^{N \FF[c(\vec\t),n,\b]} \ , \\
&\FF[c(\vec\t),n,\b] = -\sum_{\vt} c(\vt) \log c(\vt)
+ \a \log \left[ 
\sum_{\vt_1 \cdots \vt_K} c(\vt_1)\cdots c(\vt_K) 
\EE(\vt_1,\cdots,\vt_K)
\right] 
\ .
\end{split}\eeq
The partition function (\ref{repZ})
can then be evaluated by a
saddle-point, and the saddle-point value of $c(\vt)$ 
is the average of the order parameter $\r(\vt|\underline{\vs})$.
For symmetry reasons we expect that $c(\vt)=c(-\vt)$ at the
saddle point so the average over the signs $(q_1,\cdots,q_K)$ in (\ref{repZ})
can be dropped setting $q_\ell\equiv 1$.

The replica symmetric {\it ansatz} amounts to choose 
\beq\label{csimm}
c(\vt) = C\left[\sum_a \t^a\right] = \int_{-\io}^\io dz R(z) \frac{e^{\frac{\b z}{2}
    \sum_a \t^a}}{[2\cosh(\b z/2)]^n} \ ,
\eeq
where the last expression is a reparametrization of $c(\vt)$ in terms of a new
function $R(z)$ thus defined, 
and which must satisfy $R(z)=R(-z)$. The normalization $\sum_{\vt} c(\vt)=1$
implies $\int dz R(z)=1$. Substituting in (\ref{repZ}) we get, in the limit $\b\to \io$, $n\to 0$, $\n=n\b$,
\begin{eqnarray} \label{effe}
\mathcal F[R(z),\nu] &=& -\int \frac{dx \ d \hat x}{2 \pi} \ e^{i
x \hat x + \frac{1}{2} \nu | \hat x| } \varphi(x) \log \varphi(x) +
\alpha \log \int_{-\infty}^{+\infty} dz_1 \dots dz_K R(z_1) \dots
R(z_K) e^{\nu \Phi(\bf z)} \ , \label{fdirnu}
\end{eqnarray}
where
\begin{eqnarray}
\varphi(x) &=& \int dz \ e^{-ixz-\frac{1}{2}\nu |z|} R(z) \ , 
\label{varphidef}
\\ \Phi({\bf
z}) &=& \max_{{\bf \sigma} \in \{-1,1\}^K} \frac{1}{2} \sum_j \left(
\sigma_j z_j - |z_j| \right) - \mathbbm 1_{\bf \sigma, \bf 1}
 = \left\{
        \begin{array}{ll}
        -\min(1,z_1,\dots,z_K) & \textrm{if \(z_j > 0 \ \ \forall j\)
        } \\ 0 & \textrm{otherwise}
        \end{array}
\right. \ .
\end{eqnarray}

\section{Saddle-point equation}
\label{app:saddle}

Differentiating (\ref{effe}) with respect to $R(z)$, with the constraint $\int
dz R(z) = 1$, we get (${\bf z} = (z,z_2,\cdots,z_K)$):
\beq\begin{split}
0 &= \frac{\delta}{\delta R(z)} \left\{ \mathcal F[R(\cdot),\nu] +
\lambda \left[ \int R(z') \ dz' -1 \right] \right\} \\
&=- \int \frac{dx d \hat x}{2\pi} \ e^{ix \hat x +
\frac{1}{2} \nu |\hat x| - ix z - \frac{1}{2} \nu |z|}
\left[1+\log \varphi(x) \right] + \frac{\alpha K}{\mathcal D[R(\cdot)]} \int_{-\infty}^{+\infty}
dz_2 \dots dz_K \ R(z_2) \dots R(z_K)
e^{\nu \Phi(\bf z)} + \lambda
  \label{sella1}
\end{split}
\eeq
where
\begin{eqnarray}
\mathcal D[R(\cdot)] &=& \int_{-\infty}^{+\infty} dz_1 \dots dz_k \
R(z_1) \dots R(z_k) e^{\nu \Phi(z_1,\cdots,z_K)}. \label{ddir}
\end{eqnarray}
The function $R(z)$ is even, $R(z)=R(-z)$:
in principle we should add a Lagrange multiplier to enforce this constraint,
however this is equivalent to consider the equation above for $z \geq 0$ only.

In the last term, using the normalization of $R(z)$ and the definition of $\Phi$ we can write
\beq\label{B3}
\begin{split}
&\int_{-\infty}^{\infty}
dz_2 \dots dz_K \ R(z_2) \dots R(z_K)
e^{\nu \Phi(\bf z)}
= 1-\frac{1}{2^{K-1}} + \int_{0}^{\infty}
dz_2 \dots dz_K \ R(z_2) \dots R(z_K)
e^{\nu \Phi(\bf z)} \\
&=  1-\frac{1}{2^{K-1}} + \int \frac{dx d\hat x}{2 \p} e^{-\n \min(z,\hat x) -
  i x \hat x}
\int_{0}^{\infty} dz_2 \dots dz_K \ R(z_2) \dots R(z_K) e^{i x \min(1,z_2,\cdots,z_K)}.
\end{split}
\eeq
Defining
\beq
Q(x) = \int_{0}^{\infty} dz_2 \dots dz_K \ R(z_2) \dots R(z_K) e^{i x
  \min(1,z_2,\cdots,z_K)} \ ,
\eeq
and using the relation 
$\min(z,\hat x) = -\frac12 \big[ |z-\hat x| - z - \hat x \big]$,
the last integral in (\ref{B3}) can be written as
\beq
\int \frac{dx d\hat x}{2 \p} e^{-\n \min(z,\hat x) -
  i x \hat x} Q(x) =
\int \frac{dx d\hat x}{2 \p} e^{-i x z + i x \hat x - \frac\n2 z + \frac\n2
  |\hat x|
} Q\left(x+\frac{i\n}2\right) = \int dx K(z,x) Q\left(x+\frac{i\n}2\right) \ ,
\eeq
having defined the kernel 
$K(z,x) = \int \frac{d\hat x}{2 \p} e^{-i x z + i x \hat x - \frac\n2 z + \frac\n2
 | \hat x | }$,
that appears also in equation (\ref{sella1}) for $z\geq 0$; note that $\int dx K(z,x)=1$.
The saddle point equation (\ref{sella1}) then becomes, for $z \geq 0$:
\beq\label{sella1bis}
0=\int dx K(z,x) \left\{ \l-1 - \log \f(x) + \frac{\a K}{\DD[R(\cdot)]} 
\left[ 1 - \frac{1}{2^{K-1}} + Q\left(x+\frac{i\n}2\right) \right] \right\} \ .
\eeq
A solution of this equation is obtained when the term into curly
brackets vanishes. Inverting Eq.~(\ref{varphidef}),
$R(z)=\int \frac{dx}{2\pi} e^{ixz + \frac{\n}{2}|z|} \f(x)$, and expressing
$\f(x)$ using (\ref{sella1bis}) we get
\beq\label{sella3}
\begin{split}
R(z) = \int_{-\infty}^{+\infty} \frac{dx}{2\pi}
\exp\left\{ ixz + \frac{\n}{2}
|z|  + \lambda-1+\frac{\alpha K}{\mathcal D[R(\cdot)]} 
\left[1-\frac{1}{2^{K-1}}  + Q\left(x+\frac{i\n}2\right)           \right] 
 \right\} \ .
\end{split}\eeq
Substituting
\begin{equation}
R(z) = \sum_{n=-\infty}^{+\infty} r_n \ \delta(z-n)
\end{equation}
in (\ref{sella3}) we obtain the coefficients:
\begin{eqnarray}
r_n &=& \frac{e^{\frac{1}{2}\nu |n|} I_n(\alpha K
B)}{\sum_{n'=-\infty}^{+\infty}e^{\frac{1}{2}\nu |n'|} I_{n'}(\alpha K
B)} \label{rn} \ , \\ B &=&
\frac{\left(\frac{1-r_0}{2}\right)^{K-1}e^{-\frac{1}{2}\nu}}{1+\left(\frac{1-r_0}{2}\right)^K
\left(e^{-\nu}-1\right)} \ , \label{B}
\end{eqnarray}
where the denominator of \(B\) is \(\mathcal D[R(\cdot)]\) from Eq. (\ref{ddir}) and $I_n(x)$ is the modified Bessel function of order $n$.

\section{Ground state energy}
\label{app:energy}

We want to compute the ground state energy
\beq\begin{split}
 \overline{e_0}(\n) & = - \frac{\partial}{\partial \nu} \mathcal
F[R(\cdot),\nu] = \frac12 \int \frac{dx d \hat
x}{2\pi} \ e^{ix \hat x + \frac{1}{2} \nu |\hat x|} \left\{
 | \hat x | \ \varphi(x) \log \varphi(x) - \left[1+\log
\varphi(x)\right] \int_{-\infty}^{+\infty} dz \
e^{-ixz-\frac{1}{2}\nu|z|} |z| R(z) \right\}
+ \nonumber \\ & - \alpha \int_{-\infty}^{+\infty} dz_1 \dots dz_K \
\frac{R(z_1) \dots R(z_K)}{\mathcal D[R(\cdot)]} \Phi({\bf z}) e^{\nu
\Phi({\bf z})} \ .
\end{split}\eeq
We can use the saddle point equations (\ref{sella1}) and (\ref{sella3}) to eliminate the integrals \(dx \ d \hat x\) and obtain:
\beq\begin{split}
& \overline{e_0}(\n) = - \int_0^\infty dz \ z R(z) + \frac{\alpha K}{4}
\int_0^\infty dz_2 \dots dz_K \ \frac{R(z_2) \dots R(z_K)}{\mathcal
D[R(\cdot)]} \min(1,z_2,\dots,z_K) + \nonumber \\ 
&+ \alpha
 \int_0^\infty dz_1 \dots dz_K \ \frac{R(z_1) \dots
R(z_K)}{\mathcal D[R(\cdot)]} \left[ \frac{K}{2} \min(1,z_2,\dots,z_K) +(1-K)\min(1,z_1,\dots,z_K)\right]
e^{-\nu \min(1,z_1,\dots.,z_K)} \ .
\end{split}\eeq
Using Eq.~(\ref{sommarn}) for $R(z)$ we have
\begin{eqnarray}
 \overline{e_0}(\n)  &=& -\sum_{n=1}^\infty n \ r_n + \frac{\alpha K}{2}
\frac{\left(\frac{1-r0}{2}\right)^{K-1}\left(\frac{1+r_0}{2}\right)}{1+\left(\frac{1-r0}{2}\right)^K
\left(e^{-\nu}-1 \right)} +\alpha \left(1-\frac{K}{2}\right)
\frac{\left(\frac{1-r0}{2}\right)^K e^{-\nu}}
{1+\left(\frac{1-r0}{2}\right)^K \left(e^{-\nu}-1 \right)}
\end{eqnarray}
and from the expressions (\ref{rn}) and (\ref{B}) for $r_n$ and $B$ we obtain
\begin{eqnarray}
 \overline{e_0}(\n)  &=& -\frac{\partial}{\partial \nu} \log \mathcal
I(\alpha K B,\nu) + \frac{\alpha K B}{2} \frac{1+r_0}{2} \
e^{\frac{1}{2}\nu} + \alpha \left(1-\frac{K}{2}\right) B \
\frac{1-r_0}{2} \ e^{-\frac{1}{2}\nu} \label{EGS} \\ \mathcal I(z,\nu)
& \equiv & \sum_{n=-\infty}^{+\infty} e^{\frac{1}{2} \nu |n|} I_n(z)
\end{eqnarray}
For $\n>0$ this sum is always converging, as can be seen from Eq. (\ref{somma}).

\subsection{The limit $\n\to\io$}

We are interested in the limit $\n\to\io$ as in this limit 
$\overline{e_0}(\n) \to
0$ as we will show.  Let us define $\ee \equiv e^{-\nu}$ and from
(\ref{B}) write:
\beq\begin{split} &G \equiv \frac{\a K B e^{\frac12 \nu}}2
= \frac{\a K}2
\frac{\left(\frac{1-r_0}{2}\right)^{K-1}}{1-\left(\frac{1-r_0}{2}\right)^K
(1-\ee)} = \frac{\a K}2
\frac{\left(\frac{1-r_0}{2}\right)^{K-1}}{1-\left(\frac{1-r_0}{2}\right)^K
} \left[ 1 - \ee
\frac{\left(\frac{1-r_0}{2}\right)^{K}}{1-\left(\frac{1-r_0}{2}\right)^K
}\right] \equiv \GG \left[ 1 - \ee \frac{2 \GG}{\a K}
\frac{1-r_0}{2} \right], \\ 
&\GG = \frac{\a K}2
\frac{\left(\frac{1-r_0}{2}\right)^{K-1}}{1-\left(\frac{1-r_0}{2}\right)^K
}.
\end{split}
\eeq 
Using the small-$z$ expansion of the Bessel functions $I_n(z)$ ($n\geq 0$) 
\beq\label{Ias} I_n(z) \sim
\frac{z^{n}}{2^{n} n!} \left[ 1 + \frac{z^2}{4(n+1)} + O(z^4) \right]
\eeq 
and $I_{-n}(z)=I_n(z)$ we have, using the identities 
\beq \sum_{n=-\io}^\io \frac{G^{|n|}}{|n|!} = 2e^G -1 \ ,
\hskip30pt G^2 \sum_{n=-\io}^\io \frac{G^{|n|}}{(|n|+1)!} = 2G e^G
- G^2 - 2G \ , \eeq that \beq \II(\a K B,\n) \sim \sum_{n=-\io}^\io
\frac{G^{|n|}}{|n|!}  \left[ 1 + \frac{\ee \, G^2}{|n|+1} + O(\ee^2)
\right] = 2 e^{G} - 1 + \ee (2G e^G - G^2 - 2G) + O(\ee^2) .
\eeq
The equation for $r_0$ is from (\ref{rn}), (\ref{Ias})
\beq\begin{split} r_0 &= \frac{I_0(2G e^{\frac12\n})}{2 e^{G} - 1+
\ee (2G e^G - G^2 - 2G) + O(\ee^2) } =\frac{ 1 + \ee G^2 +
O(\ee^2) }{2 e^{G} - 1+ \ee (2G e^G - G^2 - 2G) + O(\ee^2) } \\
&= \frac{1}{2e^{\GG}-1} \left\{ 1 + \ee \GG^2 +
\frac{\ee}{2e^{\GG}-1} \left[ 2 e^{\GG} \frac{2\GG}{\a K}
\frac{1-r_0}{2} - 2\GG e^{\GG} + \GG^2 + 2\GG \right] + O(\ee^2)
\right\} \\ &= F_0(r_0) + \ee F_1(r_0) + O(\ee^2) \ .
\end{split}
\eeq 
The solution to the previous equation is 
\beq\label{rho0}
\begin{split}
&r_0 = \r_0 + \ee \r_1 \ , \\ &\r_0 = \frac{1}{2e^{\GG(\r_0)}-1} \ ,
\\ &\r_1 = \frac{F_1(\r_0)}{1-F_0'(\r_0)}.
\end{split}
\eeq 
To write the energy we need to compute 
\beq \sum_{n=1}^\io n
e^{\frac12 n \n} I_n(\a K B) = \sum_{n=1}^\io \frac{G^{n}}{(n-1)!}
\left[ 1 + \frac{\ee \, G^2}{n+1} + O(\ee^2) \right] = G e^G + \ee
G (1-e^G + G e^G) + O(\ee^2) .
\eeq 
The energy (\ref{EGS}) is then given by, neglecting $O(\ee^2)$: 
\beq \overline{e_0}(\n) =- \frac{G e^G + \ee G
(1-e^G + G e^G)}{2 e^{G} - 1 + \ee (2G e^G - G^2 - 2G)} + G
\frac{1+r_0}{2} + \ee \left(\frac{2}{K} -1 \right) G \frac{1-r_0}{2}.
\eeq
Given that $\overline{e_0}(\n) = -\partial_\nu F$ and that $r_0$ is the solution
of $\partial_{r_0} F =0$ the term $\ee \r_1$ in $r_0$ should not
contribute to $\overline{e_0}(\n)$ at first order in $\ee$. Then we can write
\beq\label{egscompleta}
\begin{split} \overline{e_0}(\n)= - \GG e^{\GG} \r_0 & \left\{ 1 - \ee
\frac{2 \GG^2}{\a K} \frac{1-\r_0}{2} \left[\frac{1}{\GG} - \r_0
\right] + \ee \left[ e^{-\GG} - 1 + \GG \right] -\ee \r_0 \left[ 2
\GG e^{\GG} - \GG^2 - 2\GG \right] \right\} \\ &+\GG \r_0
e^{\GG} \left[ 1 - \ee \frac{2 \GG}{\a K} \frac{1-\r_0}{2} \right] +
\ee \left(\frac{2}{K} -1 \right) \GG \frac{1-\r_0}{2} \\ = \ee \GG
e^{\GG} \r_0 & \left\{ -\frac{ \GG^2 \r_0(1-\r_0)}{\a K} -\left[
e^{-\GG} - 1 + \GG \right] +\r_0 \left[ 2 \GG e^{\GG} - \GG^2 -
2\GG \right] + \frac1{\r_0 e^{\GG}} \left(\frac{2}{K} -1 \right)
\frac{1-\r_0}{2} \right\}
\end{split}
\eeq and $\overline{e_0}(\n) \sim e^{-\n}$ for large $\n$.  
The latter expression is complicated, but it simplifies
considerably in the limit $\a \to \io$.

\section{Free energy of the RS solution}
\label{app:free}

Finally, we can compute the free energy corresponding to the solution
to (\ref{rn}). We begin by calculating
\begin{eqnarray}
\varphi(x) = \int dz \ e^{-ixz-\frac{1}{2}\nu |z|}
\sum_{n=-\infty}^{\infty} \frac{e^{\frac{1}{2}\nu |n|} I_n(\alpha K
B)}{\mathcal I(\alpha K B, \nu)} \ \delta(z-n) = \frac{e^{\alpha K B \cos x}}{\mathcal
I(\alpha K B,\nu)} \ .
\end{eqnarray}
Then using $\varphi(x) \log \varphi(x) =
\left[\frac{\partial}{\partial p} \ \varphi(x)^p \right]_{p=1}$
we rewrite the first term in (\ref{fdirnu}) as
\begin{eqnarray}
\int \frac{dx \ d\hat x}{2 \pi} e^{ix\hat x +\frac{1}{2}\nu |\hat x|}
\varphi(x) \log \varphi(x) &=& \left. \frac{\partial}{\partial p}
\right|_{p=1} \int \frac{dx \ d\hat x}{2 \pi} e^{ix\hat x
+\frac{1}{2}\nu |\hat x|} \varphi(x) ^ p \nonumber \\ &=&
\left. \frac{\partial}{\partial p} \right|_{p=1} \int \frac{dx \ d\hat
x}{2 \pi} e^{ix\hat x +\frac{1}{2}\nu |\hat x|} \times \frac{e^{p \
\alpha K B \cos x}}{\mathcal I(\alpha K B)^p} \nonumber \\ &=&
\left. \frac{\partial}{\partial p} \right|_{p=1} \int d \hat x \
\frac{e^{\frac{1}{2}\nu |\hat x|}}{\mathcal I(\alpha K B)^p} \int
\frac{dx}{2\pi} \ e^{p \ \alpha K B \cos x} \cos(\hat x x)
\end{eqnarray}
where all integrals are between $-\infty$ and $+\infty$. The $dx$
integral is of the form \beq f(\hat x) = \int_{-\infty}^{+\infty}
\frac{dx}{2\pi} \ \cos(\hat x x) \ \psi(\cos x) =
\sum_{n=-\infty}^{\infty} f_n \ \delta(n-\hat x) \eeq where \beq f_n =
\int_0^{2\pi} \frac{dt}{2\pi} \ e^{itn} \psi(\cos t) = \frac{1}{\pi}
\int_0^\pi dt \ e^{p\ \alpha K B \cos t} \cos(n t) = I_n(p \ \alpha K
B).
\eeq 
In this way, we obtain for the double integral $dx \ d\hat x$
\begin{eqnarray}
\left. \frac{\partial}{\partial p} \right|_{p=1}
\frac{\sum_{n=-\infty}^\infty e^{\frac{1}{2}\nu |n|} I_n(p \ \alpha K
B)}{\mathcal I(\alpha K B,\nu)^p} = \frac{\alpha K B \mathcal
I^{(1,0)}(\alpha K B,\nu)}{\mathcal I(\alpha K B,\nu)} - \log \mathcal
I(\alpha K B,\nu).
\end{eqnarray}
The energy term in the free energy is just $\alpha \log \mathcal
D[R(.)]$, and we obtain \beq \mathcal F(\nu) = -\alpha K B \
\frac{\mathcal I^{(1,0)}(\alpha K B,\nu)}{\mathcal I(\alpha K B,\nu)}
+ \log \mathcal I(\alpha K B, \nu) + \alpha \log \left[ 1 +
\left(\frac{1-r_0}{2}\right)^K \left( e^{-\nu}-1 \right) \right]. \eeq
The sums can be expressed in terms of fast-converging series (for
$\nu>0$):
\begin{eqnarray}
\mathcal I(z, \nu) &=& 2 e^{z \cosh \left(\frac{1}{2}\nu\right)} -
I_0(z) - 2\sum_{n=1}^\infty e^{-\frac{1}{2}\nu n} I_n(z), \label{somma} \\ \mathcal
I^{(1,0)}(z,\nu) &=& 2 \cosh \left( \frac{1}{2} \nu \right) e^{z \cosh
\left(\frac{1}{2}\nu\right)}- e^{-\frac{1}{2}\nu}I_0(z) - 2 \cosh
\left( \frac{1}{2} \nu \right) \sum_{n=1}^\infty e^{-\frac{1}{2}\nu n}
I_n(z).
\end{eqnarray}

\section{Eigenvalues of the stability matrix in the RS solution}
\label{app:stability}

Differentiation of the free energy (\ref{repZ}) gives
\beq \label{Mst}
\begin{split}
M_{\vs\vt} \equiv \frac{\partial^2 \FF}{\partial c(\vs)\partial c(\vt)}&= -\frac{1}{c(\vs)} \d_{\vs\vt}
+ \frac{\a K (K-1) \sum_{\vs_3 \cdots \vs_K} c(\vs_3) \cdots c(\vs_K) 
\EE(\vs,\vt,\vs_3,\cdots,\vs_K)}{\sum_{\vs_1 \cdots \vs_K} c(\vs_1) \cdots c(\vs_K) 
\EE(\vs_1,\cdots,\vs_K)} \\
&- \frac{\a K^2 \sum_{\vs_2 \cdots \vs_K} c(\vs_2) \cdots c(\vs_K) 
\EE(\vs,\vs_2,\cdots,\vs_K) \sum_{\vs'_2 \cdots \vs'_K} c(\vs'_2) \cdots c(\vs'_K) 
\EE(\vt,\vs'_2,\cdots,\vs'_K)}{\left[\sum_{\vs_1 \cdots \vs_K} c(\vs_1) \cdots c(\vs_K) 
\EE(\vs_1,\cdots,\vs_K)\right]^2}
 \ ,
\end{split}\eeq
with $\EE$ as in (\ref{epsilonG}).
The function $c(\vs)$ can be computed at the saddle point using
(\ref{csimm}), (\ref{sommarn}) and (\ref{Poisson0}); defining $s = \frac1n
\sum_a \s_a$ we have
\beq
c(\vs) = \frac{1}{2e^{\GG}-1} \left[ e^{\GG e^{-\frac{\n (1-s)}2}} + 
e^{\GG e^{\frac{-\n (1+s)}2}} -1 \right] 
\sim  \frac{1}{2e^{\GG}-1}
 \left\{ e^{\GG} \d[|s|,1] + (1- \d[|s|,1]) \right\}
\ ,
\eeq
where the last equality holds for $\n\to\io$. For large $\a$, $\GG$ becomes large and the expression for $c(\vs)$ further simplifies:
\beq
c(\vs)=\frac{1}{2} \d(|s|,1) + O(e^{-\a}).
\eeq
This allows a straightforward calculation of the sums appearing in (\ref{Mst}). In order to do this, we observe that $\EE(\vs_1,\dots,\vs_K)$, defined in equation (\ref{epsilonG}), is equal (in the limit $\b \to \infty$) to $1/2^K$ times the number of vectors in $\{-1,1\}^K$ that are not equal to any of the columns of the matrix whose rows are the vectors $\vs_1, \dots, \vs_K$. Then, to $o(1)$ in $\a$:
\begin{eqnarray}
\DD_K & \equiv & \sum_{\vs_1,\dots,\vs_K} c(\vs_1) \dots c(\vs_K) \ \EE(\vs_1,\dots,\vs_K) \\
&=& 2^K \times \frac{1}{2^K} \times \frac{1}{2^K}(2^K-1) = 1 - \frac{1}{2^K}
\end{eqnarray}
since the only terms that contribute to the sum are those with $\vs_i = (1,1,\dots,1)$ or $(-1,-1,\dots,-1)$, and the corresponding matrices have all the columns equal (so that there are $2^K-1$ vectors that are not equal to any column).
In the same way we obtain
\begin{eqnarray}
\DD_{K-1}(\vs) & \equiv & \sum_{\vs_2,\dots,\vs_K} c(\vs_2) \dots c(\vs_K) \ \EE(\vs,\vs_2,\dots,\vs_K) \\
&=& 2^{K-1} \times \frac{1}{2^{K-1}} \times \frac{1}{2^K} \left[ 2^K-(2-\d(|s|,1)) \right] = 1-\frac{1}{2^{K-1}}+\frac{\d(|s|,1)}{2^K}
\end{eqnarray}
since if $|s|=1$ all the columns are equal, while if $|s|<1$ there will be two different column types, and
\begin{eqnarray}
\DD_{K-2}(\vs,\vt) & \equiv & \sum_{\vs_3,\dots,\vs_K} c(\vs_3) \dots c(\vs_K) \ \EE(\vs,\vt,\vs_3,\dots,\vs_K) \\
&=& 2^{K-2} \times \frac{1}{2^{K-2}} \times \frac{1}{2^K} \left[ 2^K-D(\vs,\vt) \right] = 1 - \frac{D(\vs,\vt)}{2^K}
\end{eqnarray}
where the function $D(\vs,\vt)$ counts the number
of different pairs, among the possible four $(-1,-1)$, $(-1,1)$,
$(1,-1)$ and $(1,1)$, that actually occur in $\{(\s^a,\t^a),\ a=1,\dots,n\}$.
It is straightforward to verify the recursion relations
\begin{eqnarray}
\DD_K = \sum_{\vs} c(\vs) \ \DD_{K-1}(\vs) \ , \\
\DD_{K-1}(\vs) = \sum_{\vt} c(\vt) \ \DD_{K-2}(\vs,\vt) \ .
\end{eqnarray}
Then we get, neglecting $e^{-O(\a)}$:
\beq\begin{split}
M_{\vs\vt} &= - \frac{2e^{\GG} \d_{\vs\vt}}{e^{\GG} \d[|s|,1] + (1-
  \d[|s|,1])} 
+\frac{\a K (K-1) [2^{K} - D(\vs,\vt) ]}{2^K-1} \\&- \frac{\a K^2 (\d[|s|,1]
  + 2^{K} -2) (\d[|t|,1]+ 2^{K} -2)
}{(2^{K}-1)^2}
 \ .
\end{split}
\eeq
This matrix is invariant under permutations of the replicas so it preserves
the symmetry of the vectors under permutations. This means that it can be
block-diagonalized in subspaces of given replica symmetry.

First we have to take into account the constraint $\sum_{\vs} c(\vs)=1$.
This can be done by considering $\FF[c(\vs)]=\FF[1-\sum_{\vs}c'(\vs),c'(\vs)]$
where $c'(\vs)=c(\vs)$ for $\vs \neq (+1,\cdots,+1) \equiv \vec 1$ and
$c'(\vs)$ has no $\vec 1$ component. Then it is
easy to show that Hessian matrix of $\FF$ with respect to $c'$ is
($\vs,\vt \neq \vec 1$):
\beq\begin{split}
M'_{\vs\vt} &= M_{\vs\vt} - M_{\vs\vec 1} - M_{\vec 1\vt} + M_{\vec 1 \vec 1}
=  - \frac{2e^{\GG}\d_{\vs\vt}}{e^{\GG} \d[|s|,1] + (1-
  \d[|s|,1])} -2 \\
&-\frac{\a K (K-1)}{2^K-1} [ 1 + D(\vs,\vt) - D(\vec 1,\vt) - D(\vs,\vec 1) ]
- \frac{\a K^2  (\d[|s|,1]-2)(\d[|t|,1]-2)}{(2^{K}-1)^2}  
\ .
\end{split}\eeq

Let us start with the non-symmetric subspaces. In these subspaces, $|s| \neq
1$ and $|t| \neq 1$, then
\beq
M'_{\vs\vt} = - 2e^{\GG}\d_{\vs\vt}
-\frac{\a K (K-1)}{2^K-1} [ 1 + D(\vs,\vt) - D(\vec 1,\vt) - D(\vs,\vec 1) ]
-2 
- \frac{4\a K^2}{(2^{K}-1)^2} \ .
\eeq
The diagonal term is $O(e^{\a})$ while the off-diagonal part is $O(\a)$. This means that even in the properly-symmetrized basis the matrix elements will have a diagonal part of $O(e^{\a})$ while the off-diagonal elements will be $O(\a)$. Then it is easy
to show (\eg in perturbation theory) that the off-diagonal terms can change the
eigenvalues at most by a quantity $O(\a 2^n)$, so it cannot change the sign of
the eigenvalues. In this space the matrix $M$ has then only negative
eigenvalues and $\FF$ has a maximum.

In the symmetric subspace we can use the same argument for all the eigenvalues but
the diagonal element corresponding to
 $\vs,\vt=-\vec 1$ which is not $O(e^\a)$.
However we can write
\beq
M'_{\vs\vt} = 
-2 e^{\GG} \d_{\vs\vt} (1-\d[|s|,1]) + V_{\vs\vt} \ , 
\eeq
and treat $V$ as a perturbation. In the dangerous subspace $\vs=\vt=-\vec 1$
where the diagonal part has zero eigenvalue, the matrix element of the
perturbation is 
\beq
V_{-\vec 1,-\vec 1} = M'_{-\vec 1,-\vec 1} = 
M_{-\vec 1,-\vec 1} - M_{-\vec 1,\vec 1} - M_{\vec 1,-\vec 1} + M_{\vec
  1,\vec 1} =
 -4 < 0 \ ,
\eeq
so the eigenvalues of $M'$ are all negative for $\a$ large enough.

\section{Solutions with non-integer fields in the $\nu\to\infty$ limit}
\label{app:nonint}

We look for a solution with rational-valued fields,
\begin{equation}
R(z) = \sum_{n=-\infty}^{+\infty} r_n \ \delta\left( z-\frac np \right) \label{sommarndemi}
\end{equation}
where $p$ is an integer $\ge 1$. As the fields are expected to be
integer-valued, the existence of such a solution would be an indication for an
instability of the replica symmetric solution.
We plug this {\it ansatz} in the self-consistent equation for $R$
(\ref{sella3}) and find a self-consistent equation for the $p$
variables $r_0,r_1,\ldots,r_{p-1}$.
\begin{equation}
\sum_{n=-\infty}^{+\infty} r_n \ \cos( x \frac np)\ e^{-\nu |n|/(2p)} =
\exp \left( \mu + \alpha K\,
\sum _{q=1}^p A_q \; \cos( x \frac qp )\; e ^{-\nu q/(2p)}
\right)
\end{equation}
which must be true for any $x$. In the above equation we have defined 
\begin{equation}
A_1 = \frac{w ^{K-1} - \left(w -r_1\right) ^{K-1}} { 1-w ^{K}}\ , \ 
A_q = \frac{(w-r_{q-1}) ^{K-1} - \left(w -r_{q}\right) ^{K-1}} { 1-w ^{K}}\
 (2\le q\le p-1), \ \ldots \ ,
A_p = \frac{\left(w -r_{p-1}\right) ^{K-1}} { 1-w ^{K}}\ , \ 
\end{equation}
where $w=(1-r_0)/2$. To calculate the constant $\mu$ we set $x=i \frac
{\nu}2$ and send $\nu\to\infty$ to obtain 
\begin{equation}
\sum_{n=-\infty}^{+\infty} r_n \ \left(\frac{1 + \delta _{n,0}}2\right) =
\exp \left( \mu + \frac 12 \; \alpha K\,
\sum _{q=1}^p A_q
\right) \ .
\end{equation}
In addition setting $x=0$ and sending $\nu\to\infty$  we have
$r_0 = e ^{\mu}$.
Combining the two equations above we obtain the self-consistent 
equation 
\begin{equation}
r_0 = \frac 1{2 \exp( \frac {\alpha}2 K \, \frac{w^{K-1}}{1-w^K} )
-1}
\end{equation}
which is the same equation as in the case of integer fields only
($q=1$). The equation for the weight of the smallest non zero field
reads
\begin{equation}\label{erre1}
r_1 = r_0 \ \frac {A_1}2 \ ,
\end{equation}
and for $r_1 \neq 0$ can be written equivalently, with $y=r_1/w$, as
\begin{equation}
\frac{y}{1-(1-y)^{K-1}} = \alpha K\; 
\left(\frac{1}2-w\right) \; \frac{w^{K-2}}{1-w^K} \ .
\end{equation}
As $r_0$ ranges from 0 to 1, $w$ ranges from 0 to $\frac 12$.
Moreover, note that $r_1 \leq w$ because $w$ is the probability of having
a positive field; then $y$ ranges from 0 to 1.
When $\alpha$ is large the r.h.s. is $\sim \gamma
e^{-\gamma}$ which is very small while the l.h.s. is larger than
$1/(K-1)$ (minimal value in $y=0$). Therefore the latter equation has no
solution and the only solution to (\ref{erre1}) is $r_1=0$.

\end{document}